\documentclass[preprint,showpacs,preprintnumbers,amsmath,amssymb]{revtex4}

\usepackage{graphicx}

\begin{document}
\title{Vortex-to-Polarization Phase Transformation
Path in Pb(ZrTi)O$_3$ Nanoparticles}
\author{Ivan Naumov and Huaxiang Fu}
\affiliation{Department of Physics,  University of Arkansas,
Fayetteville, AR 72701, USA}

\date{\today}

\begin{abstract}
Phase transformation in finite-size ferroelectrics is of
fundamental relevance for understanding collective behaviors and
balance of competing interactions in low-dimensional systems. We
report a first-principles effective Hamiltonian study of
vortex-to-polarization transformation in
Pb(Zr$_{0.5}$Ti$_{0.5}$)O$_3$ nanoparticles, caused by homogeneous
electric fields normal to the vortex plane. The transformation is
shown to (1) follow an unusual {\it macroscopic} path that is
symmetry non-conforming and characterized by the occurrence of a
previously unknown structure as the bridging phase; (2) lead to
the discovery of a striking collective phenomenon, revealing how
ferroelectric vortex is annihilated {\it microscopically}.
Interactions underlying these behaviors are discussed.
\end{abstract}


\maketitle

Vortex is a circularly ordered structure phase of broad interest,
which has been found in ferromagnets (FM) \cite{Shinjo},
superconductors \cite{Abrikosov}, and Bose-Einstein condensated
atoms \cite{Madison}. Recently, the vortex phase formed by {\it
electric} dipoles, as hinted by Ginzburg {\it et al.} when
exploring the possibility of superdiamagnetism \cite{Ginzburg},
was revealed to exist in ferroelectric (FE)
nanoparticles.\cite{Naumov} It was shown \cite{Naumov} that the
macroscopic toroid moment undergoes a spontaneous transition from
being null at high temperature to nonzero at low temperature,
acting as the order parameter of the vortex phase. In FEs, another
ordered structure of general concern is the ferroelectric phase
with an {\it out-of-plane} polarization\cite{Fong}. Transformation
between these two ordered structures in FE nanoparticles, by
electric fields perpendicular to the vortex plane, is a phase
transition of fundamental importance, since (i) Phase
transformation in finite-size systems has been long known as a
fundamental subject~\cite{Landau}; (ii) The vortex-to-polarization
transformation involves two phases of completely different order
parameters (namely, toroid moment in vortex phase versus
polarization in ferroelectric phase), and is a critical example to
study collective mechanism in terms of how order parameter
nucleates and/or disappears {\it within a single domain}; (iii)
Participation of external electric fields during transformation
alters and re-organizes the delicate balance among competing
interactions inside FEs \cite{Cohen}, and may lead to new
structural phases that are not possible in the zero-field case;
(iv) It concerns the interaction between FE {\it vortex} and
electric fields. Vortex, as a whole, does not interact with
electric fields because of vanished net polarization; but
individual dipoles do, which explains why collective behavior
during the vortex transformation is interesting.

Technologically, the FE vortex phase and the polarization phase
both are of considerable significance.\cite{Scott,Lines} The
former promises to increase the storage density of nonvolatile
ferroelectric random access memories by five orders of
magnitude\cite{Naumov}, while FEs in polarization phase exhibit
large strain and electromechanical response, which is of
widespread use in piezoelectric transducers and
actuators.\cite{Lines,Uchino} Nowadays, FE particles\cite{Brien},
wires\cite{Yun}, and nanotubes\cite{Morrison} were all made
possible in experiments.

Despite the fact that vortices exist both in FE and FM particles,
they are different in key aspects arising from the profound
difference between electrostatic and magnetostatic interactions.
First, the demagnetizing energy in FM particles is 10$^3$ times
smaller than the depolarizing counterpart in FE particles of the
same size, leading to the fact that for particles less than 10nm
the dominant interaction in FMs is the short-range (SR) exchange
interaction which favors magnetization, much like in
bulk\cite{Jona,Hubert}. As size increases, the demagnetization
energy (scaling approximately as N$^\frac{5}{3}$ where N is the
number of dipoles in particle) becomes more important than the SR
exchange energy (scaling as N), thereby vortex turns favorable
until domains start to play a role. Indeed, the minimum threshold
size for magnetic Ni$_{80}$Fe$_{14}$Mo$_5$ particles to adopt a
vortex structure is experimentally found as 100 nm in diameter
(particles less than this size exhibit instead
magnetization).\cite{Cowburn} This is in sharp contrast to the
situation in FE nanoparticles where the depolarizing energy is
dominant and which exhibit a vortex phase as small as 4
nm.\cite{Naumov} The balance between LR and SR interactions is
thus profoundly different in two materials. Furthermore, because
of the large depolarizing energy, the shape and anisotropy are
insignificant for FE particles when size is small ($<10$ nm). In
fact, FE vortices have been found in particles of
cylindrical\cite{Naumov}, cubic\cite{Fu}, and rectangular shapes.
As a result of these differences, the knowledge of transforming
magnetic vortex is not suitable for FE nanoparticles.

Here we perform {\it ab-initio} based simulations on
vortex-to-polarization phase transformation, induced by
homogeneous electric fields normal to the vortex plane, in
nanoparticles made of technologically important
Pb(Zr$_{0.5}$Ti$_{0.5}$)O$_3$ (PZT) solid solution. These studies
lead to (1) the discovery of a critical new structure that bridges
the transformation between two phases of unlike order parameters,
(2) an unusual collective mechanism showing how FE vortex is
annihilated, (3) the existence of a previously unknown hysteresis,
and (4) an innovative approach for reading FE toroid moment.
Neither the phase transformation path reported here, nor the
unusual dipole behaviors associated with this path, occur for {\it
inplane} electric fields (which cause only dipole
flipping~\cite{Fu}). Further, inplane fields do not produce
out-of-plane polarization that is most utilized in practice.

We use first-principles derived effective Hamiltonian
\cite{Zhong,Bellaiche} and Monte Carlo (MC) simulations to
determine dipole configuration. Calculations are performed for PZT
nanostructures of cylindrical shape, with diameter {\it d} ranging
from 9 to 25 and height {\it h}=14 (both {\it d} and {\it h} are
in units of bulk pseudo-cubic lattice parameter $a$=4.0{\AA}). The
crystallographic [001] direction is chosen as the cylindrical {\it
z}-axis, with the [100] and [010] directions being the {\it x} and
{\it y} axes. Initial dipole configuration of vortex is obtained
from annealing simulations with temperature decreased in small
step. Electric field of varied strength is applied along the +{\it
z} direction at a fixed temperature of 64K, and is coupled with
dipole {\bf p}$_i$ at cell {\it i} by -$\sum _i {\bf E}\cdot {\bf
p}_i$ as described in Ref.\onlinecite{Garcia}. Dipole-dipole
interaction in nanoparticles is handled in real space.\cite{Fu}
Typically 10000 MC sweeps are used to simulate dipole responses at
each field. Outcome of simulation is the soft-mode vector field
\{{\bf u}$_i$\} (which is directly proportional to local dipole
{\bf p}$_i$), homogeneous strain \{$\eta _i$\}, and inhomogeneous
cell shape.

\begin{figure}
\centering
\includegraphics[scale=1.0]{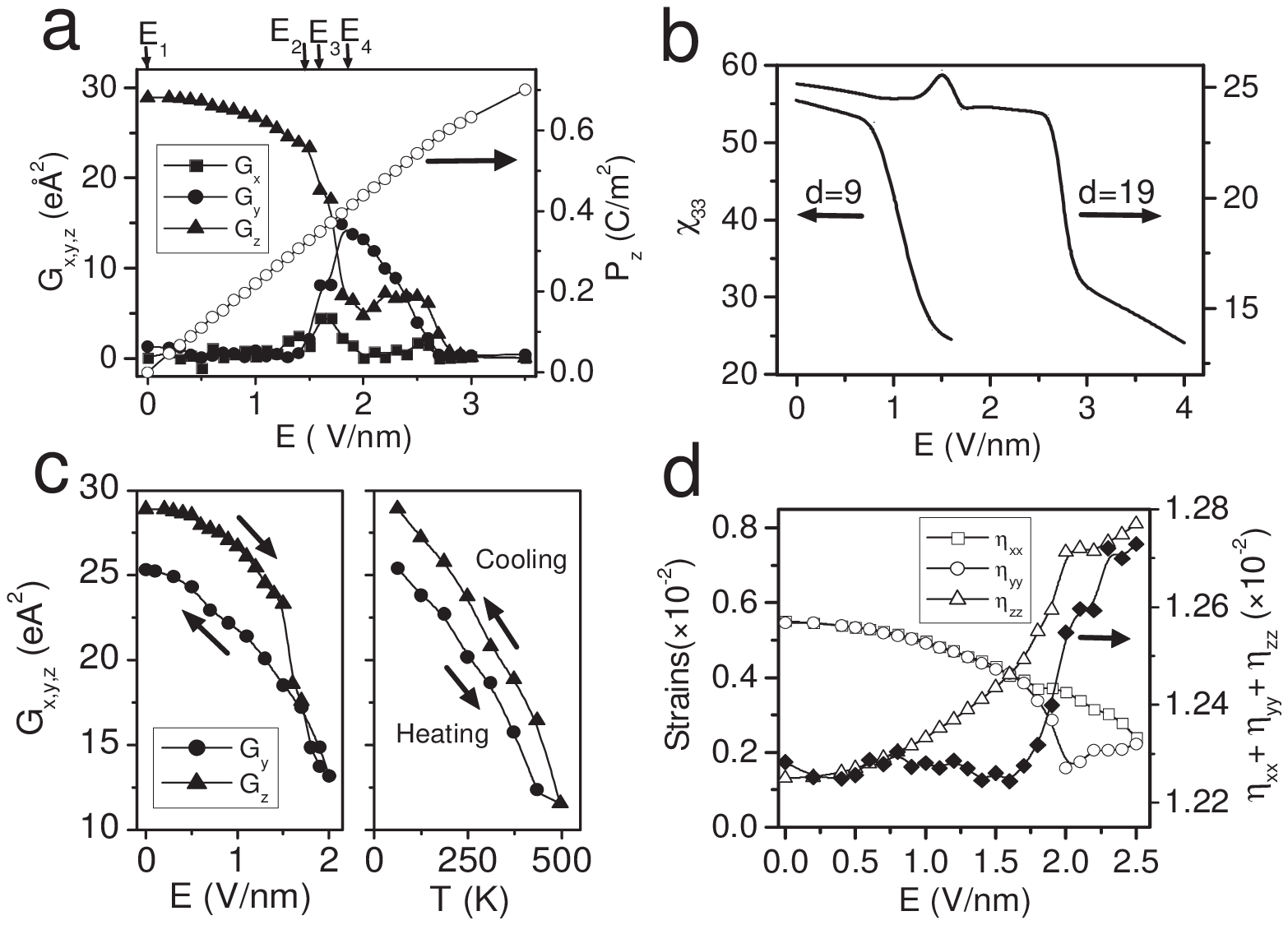}
\caption{(a) Toroid moment {\bf G} (using the left vertical axis)
and net polarization P$_z$ (using the right vertical axis) in a
d=19 disk. Arrows on the top horizontal axis indicate the selected
electric fields for which dipole configurations are analyzed. (b)
Dielectric $\chi_{33}$ susceptibility in the d=9 disk (using the
left vertical axis) and in the d=19 disk (using the right vertical
axis). (c) Left: hysteresis of toroid moment caused by increasing
and then decreasing the electric field in d=19 disk; Right:
transformation of the G$_y$ vortex into a G$_z$ vortex by heating
the system to 500K and then cooling down. (d) Strain components
$\eta_{xx}$, $\eta_{yy}$ and $\eta_{zz}$ (using the left vertical
axis) and volume expansion $\eta_{xx}+\eta_{yy}+\eta_{zz}$ (filled
symbols, using the right vertical axis) in the d=19 disk. The
horizontal axis in (a), (b), and (d) is the strength of electric
field.} \label{Fres}
\end{figure}

Figure \ref{Fres}a depicts the collective behaviors of toroid
moment, ${\bf G}=\frac{1}{2{\rm N_c}}\sum _i {\bf r}_i\times {\bf
p}_i$, and net polarization, ${\bf P}=\frac{1}{{\rm N_c}\Omega
}\sum _i {\bf p}_i$ (where N$_{\rm c}$ is the number of bulk cells
and $\Omega $ is the cell volume), that develop in a d=19 nanodisk
as the strength of electric field varies. When electric field is
small and below a critical value E$_{c,1}=1.5$V/nm, the disk shows
only a G$_z$ toroid component while G$_x$ and G$_y$ are null. The
system in the ${\rm E}\le {\rm E}_{c,1}$ field region thus retains
the same macroscopic toroid symmetry as in zero field (This
structure phase will be denoted as phase I hereafter). As the
field reaches E$_{c,1}$, the G$_z$ moment declines only slightly
as compared to the initial zero-field value, and meanwhile, a net
polarization of 0.3C/m$^2$ develops. In fact, this magnitude of
polarization is large and comparable to the value found in bulk
BaTiO$_3$.\cite{Lines,Garcia} The vortex response of phase I is
thus characterized by the coexistence of strong toroid moment and
large polarization, {\it both pointing along the cylindrical
z-axis}. We further numerically find that the toroid moment in
phase I responds to the {\bf E} field by accurately following a
quadratic scaling law as ${\rm G_z} (E)={\rm G}_{0}-\sigma {\rm
E}^{2}$, where ${\rm G}_{0}$ is the zero-field G$_z$ moment and
the $\sigma $ coefficient is determined to be 285.5
e{\AA}$^4$/V$^2$ for the d=19 disk.

The system behaves in a markedly different fashion as the electric
field exceeds E$_{c,1}$, manifested in Fig.\ref{Fres}a by the
dramatic occurrence of a nonzero G$_y$ component and
simultaneously a sharp decline of the G$_z$. Being perpendicular
to the initial G$_z$ moment, the appearance of the G$_y$ moment
deviates the system from continuing to possess macroscopic
cylindrical symmetry, and the resulting symmetry-broken new
structure is to be denoted as phase II. Thus for the first time
the ferroelectric vortex, under the application of a uniform
perpendicular field, is discovered to be able to self-organize to
generate a {\it lateral} toroid moment. Interestingly, this
behavior has never been reported for ferromagnetic vortices (to
the best of our knowledge). We have also performed calculations
for pure PbTiO$_3$ nanodisks and found that the G$_y$ toroid
moment robustly occurs, revealing that the existence of phase II
is not related to the composition fluctuation in PZT. Assuming
that dipoles respond by rotating collectively toward the field
direction, one would anticipate that the drastic decrease of the
G$_z$ moment shall be accompanied by a sharp rise in the
polarization. Surprisingly, the net P$_{z}$ polarization in
Fig.\ref{Fres}a apparently does not feel the drastic variation of
the G$_z$, and remains, to the naked eye, fairly smooth. As the
electric field exceeds a second critical value E$_{c,2}$=2.8V/nm,
all {\bf G} components vanish and the system becomes a phase of
pure polarization. Low-symmetry phase II thus acts as the key
intermediate state for bridging the transformation from phase I
(of the same cylindrical symmetry as the initial vortex) to the
destination phase of uniform polarization (that also has
cylindrical symmetry).

The evolution of toroid moment during the vortex transformation is
also reflected in dielectric $\chi _{33}$ susceptibility
(Fig.\ref{Fres}b), showing a noticeable hump at E$_{c,1}$ field.
Furthermore, near the critical E$_{c,2}$ field where vortex
collapses, the $\chi_{33}$ coefficient exhibits a sharp decline,
demonstrating that dipoles in vortex phase generate a notably {\it
larger} dielectric response than dipoles in ferroelectric phase
do.

\begin{figure}
\centering
\includegraphics[scale=0.65]{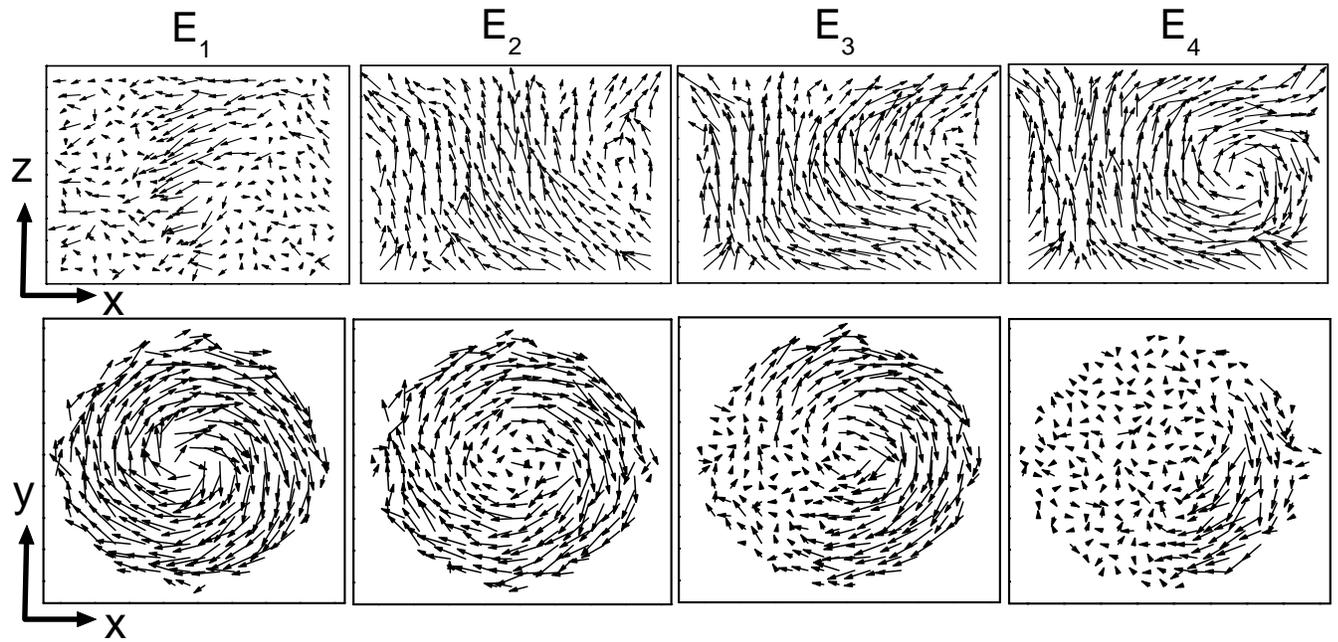}
\caption{Configurations of local dipoles on the central {\it xy}
cross section (lower panel) and on the central {\it xz} cross
section (upper panel) in the d=19 disk, at selected fields E$_1$,
E$_2$, E$_3$ and E$_4$ as marked by the arrows in Fig.\ref{Fres}a.
The magnitude of each dipole is enlarged for clarity.}
\label{Fpat}
\end{figure}

We now present {\it microscopic} understanding of the puzzling
collective behaviors that FE vortex displays in Fig.\ref{Fres}a
along the transformation path. Fig.\ref{Fpat} shows snapshots of
the dipole configurations corresponding to four selected fields
E$_i$ (i=1$\sim $4) labelled in Fig.\ref{Fres}a. Comparison of the
dipole patterns at E$_1$ and E$_2$ fields shows that the vortex
response in phase I is characterized by collective rotation of
dipoles towards the field direction. Note that at E$_2$ field the
vortex pattern on the {\it xy} plane maintains a cylindrical
symmetry. The rotation is thus symmetry conforming in the sense
that dipoles within the same distance from the central axis
respond equivalently. Another important conclusion from
simulations concerns where the polarization nucleates, since the
phase transition involve two different order parameters. We find
that the {\it z}-axis polarization is nucleated at the center of
the FE vortex. In other words, formation of polarization begins
with and is initiated by the rotation of those dipoles near the
cylindrical axis, which is consistent with the explanation that a
large strain exists in the vortex center and the rotation reduces
this strain energy.

As the field changes slightly from E$_2$ to E$_3$, two striking
phenomena occur: (1) The dipole components in the {\it xy} plane
cease to exhibit cylindrical symmetry (see the E$_3$ pattern,
lower panel, Fig.\ref{Fpat}). In fact, the inplane components
disappear remarkably for dipoles within a certain, but not across
the entire 360$^0$ degree of, azimuthal angle. (2) A {\it lateral}
G$_y$ vortex --- for which the corresponding toroid moment points
at the {\it y}-axis --- starts to nucleate near the right-side
surface on the {\it xz} plane (see the E$_3$ pattern, upper panel
of Fig.\ref{Fpat}). Note that this G$_y$ vortex forms neither at
the cylindrical axis, nor at the same side as the azimuthal angle
where inplane dipole components are annihilated, in order to avoid
the large strain energy when vortex develops.

Further increase of the electric field from E$_3$ to E$_4$ leads
to the widening of the annihilation angle (the E$_4$ pattern,
lower panel, Fig.\ref{Fpat}) and full development of the lateral
G$_y$ vortex. The system at the E$_4$ field is thus characterized
by formation of a vortex-free ferroelectric region on one side
---and a vortex region with perpendicular G$_y$ moment on the
other side (the E$_4$ pattern, upper panel, Fig.\ref{Fpat}).
Interestingly, despite that formation of the lateral G$_y$ vortex
forces some dipoles to point {\it opposite} to the field direction
between E$_3$ and E$_4$ fields, the net P$_z$ polarization
nonetheless remains increasing as seen in Fig.\ref{Fres}a,
consistent with the Le Chatelier's principle\cite{Landau}.
Finally, when electric field continues to increase above E$_4$,
the ferroelectric region expands by extruding the center of the
G$_y$ vortex toward the right-side surface, and this G$_y$ vortex
eventually disappears at critical E$_{c,2}$ field.

Though complex, the dipole behaviors in Fig.\ref{Fpat} have a
simple explanation, namely they result from the competition
between the applied electric field and depolarizing field. For
phase II at $E_{c,1}<E<E_{c,2}$, these two fields becomes
comparable, and the nanodisk desires to reduce the {\it z}-axis
depolarizing field by forming lateral G$_y$ vortex. Indeed, as
shown in Fig.\ref{Fpat} for the system at E$_4$ field, the
depolarizing field in phase II is mainly confined to the
ferroelectric region while being reduced in the vortex region.
Meanwhile, to enhance the interaction between polarization and the
external field, those dipoles {\it in the ferroelectric region}
favor to rotate toward the {\it z}-axis more than the dipoles in
the G$_y$ vortex region, leading to the collective phenomenon that
the inplane G$_z$ vortex is annihilated without maintaining the
azimuthal symmetry.

The presence of phase II (with toroid moment rotated by 90$^0$
with respect to the initial G$_z$ vortex) raises a question of
what may happen when one starts with this phase (e.g., at E=2V/nm
in Fig.\ref{Fres}a) and then decreases the external field. Our
simulations reveal that, regardless of whether the field is
gradually reduced or suddenly switched off, the system in phase II
does not transform back to the initial state of G$_z$ moment, and
instead is trapped in the G$_y$ vortex state. More specifically,
as the field is reduced, the G$_y$ vortex at the E$_4$ field in
Fig.\ref{Fpat} {\it grows} by moving its core toward the
cylindrical axis (not toward the right-side surface), resulting in
a pure vortex state with a G$_y$ moment and {\bf P}=0. This leads
to a hysteresis as shown in the left plot of Fig.\ref{Fres}c,
which is interesting in the sense that (i) the hysteresis is
caused by toroid moment, not polarization; (ii) it exists in a
single particle of nanometer size; (iii) during the hysteresis the
toroid moment is rotated rather than switched to the opposite. We
further find that the G$_y$ and G$_z$ phases at zero field in
Fig.\ref{Fres}c are very close in energy; the latter is lower by
$\sim$1 meV per 5-atom unit cell. Trapping of the system in the
G$_y$ state also suggests that this state is stable and surrounded
by energy barrier. To confirm this, the system of the G$_y$ phase
is heated at zero field to a chosen temperature $\tilde{T}$ and
then cooled down to 64K. We found that only when $\tilde{T}$ is
above 500K the G$_y$ phase is able to overcome the barrier and
becomes the G$_z$ phase (see the right plot of Fig.\ref{Fres}c).

Our analysis shows that, in addition to the reduction of the
depolarizing field, there is another factor that facilitates the
transformation from phase I into phase II, that is, the
interaction between local mode and strain. Fig.\ref{Fres}d depicts
the strain components at different fields. At zero field the
lattice of the vortex state is pseudo-tetragonal with $c/a$ ratio
less than 1, i.e., with strain components
$\eta_{xx}=\eta_{yy}>\eta_{zz}$, because all dipoles are lying in
the {\it xy} plane. As the field increases, the $c/a$ rises as a
result of the polarization-strain coupling. Notably, the field at
which $c/a$ becomes 1 (i.e., the system becomes pseudocubic) is
very close to the critical E$_{c,1}$ field where phase I is
transformed into phase II. The mode-strain coupling\cite{Zhong}
$-\sum_{i}|B| ( \eta_{xx}u_{ix}^{2}+\eta_{yy}u_{iy}^{2}+
\eta_{zz}u_{iz}^{2})$ (where B is the coupling strength) advances
the transition into phase II largely due to the increase in atomic
volume, $\Delta \Omega/\Omega_{0} =\eta_{xx}+\eta_{yy}
+\eta_{zz}$, which is nearly a constant in phase I and rises
sharply for E$>$E$_{c,1}$ (Fig.\ref{Fres}d).

We next address how the diameter of nanodisk may influence the
vortex transformation. Interestingly, we find that there is a
critical size d$_c$=17, below which the transformation path turns
out to be different. Simulation results of a d=9 disk (not shown
here) reveal that the vortex undergoes a continuous transformation
into a single domain ferroelectric phase, without the bridging
phase II. The dielectric $\chi_{33}$ coefficient for this disk at
zero field is determined to be 55 (Fig.\ref{Fres}b) and is
considerably larger than the value of 25 in the d=19 disk, showing
that the same electric field is able to induce a much larger
polarization in {\it smaller} disks. The critical E$_{c,2}$ field
in the d=9 disk decreases to 1.3V/nm. One main difference between
d=9 and d=19 nanodisks is the depolarizing effect which is small
in the former and thus allows a symmetry conforming
transformation.

Finally we point out that the predicted FE-vortex responses have
important technological implication. The quadratic law, $G_z
(E)=G_{0}-\sigma E^{2}$, tells us that, when a FE nanodisk is
exposed to an alternating field $\mathbf {E}(t)=\mathbf
{E}_{0}\cos\, \omega t$, the toroid moment $\mathbf{G}(t)$ will
respond with a double frequency $2\omega$, and its radiation field
can be separated from the field of the vibrating polarization
which responds only with $\omega$. The signal with $2\omega$
frequency may further tell us whether it is associated with moment
$\mathbf {G}$ or $-\mathbf {G}$, since the latter field is phase
shifted by $\pi$.\cite{Dubovik} This may thus open a novel
approach by using electromagnetic fields of pulse lasers to probe
and/or read FE vortex state. Compared to mechanical approach using
piezoelectric force microscope tips, the optic approach is fast
and can be performed simultaneously in a large quantity.

In summary, (i) the transformation between vortex and
ferroelectric phases in FE particles is predicted to follow an
unusual and symmetry non-conforming path, which is a macroscopic
manifestation of the delicate competition of microscopic
interactions. A new structure phase, with rotated vortex moment
and coexisting polarization, was revealed as the intermediate
phase bridging the transformation. (ii) We discovered a striking
collective phenomenon leading to the annihilation of FE vortex,
namely that the vortex does not disappear shell-by-shell, but in a
peculiar azimuthal annihilation mode. (iii) The center of FE
vortex plays a crucial role for the nucleation of polarization.
(iv) The existence of an interesting hysteresis, caused by toroid
moment, may act as a previously unknown channel for energy
dissipation and dielectric loss. (v) Our simulations further
suggest an innovative optic approach for reading/probing toroid
moment.

This work was supported by the National Science Foundation and
Office of Naval Research. The computing facilities were provided
by the Center for Piezoelectrics by Design.

\newpage

\end{document}